**Comment on "Discussions on common errors in analyzing sea level accelerations, solar trends and global warming" by Scafetta (2013a,b).**
*R.E. Benestad*

The regression analysis discussed in Benestad and Schmidt (2009; BS09) is misrepresented in the papers Scafetta (2013a,b). In the abstract of BS09, it is stated that "*We demonstrate that naive application of linear analytical methods such as regression gives nonrobust results*". The paper iterates this point further "*The regression analysis ... should in this context be regarded as a naive approach that is prone to yielding biased results, and we caution against using such techniques without a critical interpretation*". To reiterate this point for the third time, the BS09 reads: "*Here we use the regression to demonstrate how spurious results may arise from colinearity and ''noise'' by examining the variability in the coefficients*".

The very same demonstration referred to above was presented in the papers Scafetta (2013a,b; S13a,b) as: "*An improper application of the multilinear regression method is found in Benestad and Schmidt (2009), indicated herein as BS09*". S13a,b further misrepresented BS09 by stating "*The first way BS09 multi-linear regression fails is mathematical. The predictors of a multilinear regression model must be sufficiently linearly independent, i.e. it should not be possible to express any predictor as a linear combination of the others*" without pointing out that this was exactly the point made in BS09 too.

While the purpose of BS09 was to show why such methods fail, S13a,b turned this around and accused BS09 for inappropriate use of this method. Hence, I think the way Scafetta refers to BS09 is dishonest, and I will recommend the readers to read the original paper and make up their own mind[1].

Moreover, S13a,b did not give BS09 the credit for using the regression coefficients as a means for evaluation of the method, or what Scafetta referred to as the 'scaling factors'. BS09 examined their values and argued that values difficult to reconcile with physics provided an additional indication of flaws.

S13a,b misrepresent BS09 by giving the impression that a multiple regression with 10 covariates was used to estimate the solar contribution to the recent warming. The regression analysis in BS09 used for comparing climate models and observations only included two co-variates. S13a,b, makes no mention of this fact, and gives a false impression that a regression with 10 covariates was used for the comparison and the conclusion of a 7% solar contribution.

The choice of boundary settings for the wavelet analysis discussed in Scafetta (2013a,b) does not affect the conclusions of BS09, and the analysis has been repeated with the same settings as in Scafetta and West (2006). Furthermore, the analysis did not only use wavelets for band-pass filtering, but also included other approaches, and the conclusions were not sensitive to the choice of filtering strategy, as explained in BS09. Moreover, BS09 argued that the problem was something else: "*taking*

---
[1] Freely available from NASA: http://pubs.giss.nasa.gov/abs/be02100q.html

*the relative magnitudes between two band-pass filtered signals, does not identify a true connection between the two"*. Scafetta and West (2006) had estimated the effect of solar variability on Earth's temperature by first band-pass filtering both over the 14.7–29.3 years (in addition to 7.3–14.7 years) respectively, and then take the ratio of the standard deviation $\sigma_{temp}/\sigma_{sun}$ for the band-pass filtered series as a measure of the response. By adopting this ratio, Scafetta and West (2006) implicitly and incorrectly assumed that no other factors were involved with time scales of 7.3-29.3 years, and that all of the temperature changes with those time scales were due to changes in the sun. The BS09 paper presented a critical analysis of a number of papers of Scafetta's.

The reason for using different types of boundary conditions in BS09 compared to Scafetta and West (2006) was insufficient information about the details of the analysis. Likewise, Scafetta (2013a,b) lacks information about how the regression model was calibrated (the time period was not clear, but it seemed to be from 1980 to 2004), which makes his claim difficult to verify (He argues that the solar contribution of the global warming between 1900 and 2000 is similar to the anthropogenic forcings). It seems that the calibration over such a short interval is likely to miss the long-term changes due to changes in the greenhouse gas concentrations, masked by the short-term fluctuations.

S13a,b also appears to make self-contradictory claims by arguing that $CO_2$ and $CH_4$ boost the effect of changes in the sun (assuming they are released from a warming caused by changes in the sun simultaneously) and simultaneously assume that the effect from anthropogenic $CO_2$ is weak. S13a,b further makes reference to "outdated hockey-stick paleaoclimatic temperature graphs" with no factual support. Both these assertions were hand-wavy arguments not supported by evidence. Contrary to the assertions made in S13a,b, the "outdated hockey-stick paleaoclimatic temperature graphs" were presented in the most recent IPCC assessment report (Solomon et al., 2007).

Scafetta (2013a,b) also makes a number of statements about the sea-level, however, this aspect is not considered here in this comment.